\documentclass[aps,showpacs,prd,twocolumn]{revtex4}
\usepackage{graphicx, epsfig}
\usepackage[T1]{fontenc}
\usepackage{amssymb}
\usepackage{amsmath}
\usepackage{latexsym}
\newcommand{\be}{\begin{equation}}
\newcommand{\ee}{\end{equation}}
\newcommand{\ben}{\begin{eqnarray}}
\newcommand{\een}{\end{eqnarray}}
\newcommand{\bes}{\begin{subequations}}
\newcommand{\ees}{\end{subequations}}
\newcommand{\bb}{\bibitem}

\begin{document}
\title{Nuclear Phenomenology: A Conceptual Proposal for High School Teaching}
\author{J.D. Dantas and M.A.M. Souza}
\email[]{emails:msouza@fisica.ufpb.br/ m.a.m.souza.ufpb@gmail.com}
\affiliation{Departamento de F\'\i sica, Universidade Federal da Para\'\i ba, Caixa Postal 5008, 58051-970 Jo\~ao Pessoa, Para\'\i ba, Brasil}

\begin{abstract}

The discovery of atomic nucleus by E. Rutherford, at the beginning of the twentieth century, was the Nuclear Physics original landmark. From then, a series of experiments in which beams of particles composed of neutrons, protons and others, brought to collide with a nucleus  in order to unravel its structure or produce artificial elements through nuclear transmutation, were triggered. With the development of experimental equipment, a number of other nuclear phenomena have been observed, such as beta decay, nuclear fission and fusion, M\"oesbauer effect, etc. In view of the global political and economic landscape and the contemporary educational trends, this work suggest alternative topics in nuclear physics that can be discussed at the conceptual level in high school teaching, where the main focus lies in the historical and technological importance of such phenomena in society.
\end{abstract}

\maketitle

\section{INTRODUCTION}

The XX century was marked by the appearance of one of the physics largest branches, the Modern Physics. The main technological progresses of the modern world are due, partly, to the appearance of the Quantum Mechanics and the Relativity Theory that were used as base to describe a series of phenomena in the atomic and nuclear scale. As example we have the superconductivity, described by the BCS theory \cite{ref1}, which has been extremely effective in the search for more efficient systems in information transmission. We also have countless progresses in Medicine, which can be in the instrumental point of view, for diagnosis with image processing equipment, such as magnetic resonance, or a clinical point of view, through radiotherapy treatments. Besides the important medical contributions, the Nuclear Physics serves as theoretical background for other knowledge areas such as Astrophysics and some ramifications of Cosmology. 

The objective of this work is to show that some themes of Nuclear Physics can be treated at a conceptual level in high school teaching. It is an effort in the search of explaining phenomena of great historical importance and practical usefulness in the modern world, according to Terrazzan proposal \cite{ref2}, that defends the Physics curriculum modernization by means of the current science development as a need to create conscious citizens capable to transform the reality. This was emphasized by Aubrecht \cite{ref3}  in the Modern Physics teaching conference in April of 1986, at FERMILAB (Fermi National Accelerator Laboratory), Batavia, Illinois, where was defended the inclusion of physics research topics in high school teaching. An interesting text about literature review dealing with the theme modern and contemporary physics in high school can be found at \cite{ref4}.

Some authors have elaborated proposals of educational methodology for themes of Modern Physics in high school teaching, like quantum mechanics  \cite{ref5} and restricted relativity \cite{ref6}. In this article, initially, we will discuss the educational motivation for the proposal of the conceptual teaching on nuclear phenomenology, whose themes, such as nuclear fission and fusion, already do part of the current menu. In subsequent sections, we will debate the phenomena conceptual structure that can be treated in High School.

\section{EDUCATIONAL MOTIVATION}

The Nuclear Physics has a prominence role in the current scientific scenario, and even in the international economic policy, concerning the uranium enrichment for military purposes or for electric energy production; what turns it into an important subject in the educational point of view. This work suggests the study of some phenomena in the nuclear scale that can be treated qualitatively in the class room at high school level, for methodological and practical finalities.

These are phenomena of high historical importance, which have applications in other knowledge areas. What is intended here is not to reformulate the course program of Modern Physics, but to make use of a theoretical background that can contribute to the student's formation in epistemological level.

Although such concepts have an extremely complex quantitative description, till certain point rooted to an empiric analysis, it is possible to turn them qualitatively simple without needing to appeal to advanced mathematical formulations, staying, however, its technical character. The student of medium level, at the beginning, would not have many difficulties in understanding the nuclear dynamics, once, when penetrating in this subject, some contents that serve as prerequisites for such understanding were already seen, partly in the disciplines of Chemistry, like the atom concept, spin, ions, bond energy, mass and atomic number, and also in the own discipline of Physics, in the topics related to Electromagnetism, Restricted Relativity and Quantum Mechanics.

The student who has studied Chemical Reactions won't have difficulties in understanding how a nuclear reaction is processed, once the notation used to describe both is identical. On the other hand, it is important that the teacher, as mediator of knowledge, is able to contextualize the contents supplied with the student's daily life; all the nuclear processes are susceptible to a practical application, and can be extended to other knowledge areas. It can stand out, among them, the geological dating with the use of isotope radioactive of $C^{14}$ or $U^{235}$, in Geography and Geology. We have the dosimetry in medical treatments by $Cs^(133)$, or the induction of genetic mutations in certain organisms in Biology, in addition to the possibility of an international socio-political analysis on the impact of the use of nuclear weapons and the handling of nuclear garbage. An experience of teaching on these last two issues was made in Holland by Eijkelhof et al \cite{ref8}, on which was included in the curriculum structure of Physics, a unit called "Nuclear Weapons and Security", where 65 percent of students agreed that the topic should be incorporated to Physics curriculum, although the administrative members of the schools have positioned themselves against inclusion

The radiological accident in Goiânia \cite{refm,refa,refb} on september $13$, $1987$, can be discussed, where $0,093$ kg of cesium chloride, a salt obtained from the cesium $137$, contained in a coated capsule with steel and lead of a radiotherapy equipment, was released contaminating $112800$ people, the most with external body contamination. Of these, $129$ people had concrete body contamination internal and external, coming to develop symptoms such as nausea, vomiting, diarrhea, etc, being just medicated. Other $49$ were hospitalized, of which $21$ needed to go through intensive treatment; from these, four did not resist and ended up dying. While cleaning the city $13,4$ tons of atomic waste were collected, separated in $1200$ boxes and $2900$ drums, which will remain as a risk source to the environment for $180$ years.

In general, the Nuclear Phenomenology is a rich topic and of great potential in the educational context. Follows as suggestion the approach of the next topics.

\section{HISTORICAl}

The description of the matter structure is an old problem which goes back to Classical Antiquity. In ancient Greece, Leucippus and Democritus defended the idea that all things were formed by bodies, to which they gave the name of Atoms. These were considered indivisible, rigid and impenetrable, equipped with an incessant movement in vacuum \cite{ref9}.

Although the ideas concerning the atomism have been abandoned and obscured by the church in the Middle Ages, at the beginning of the nineteenth century, the hypotheses on the atomic constitution of matter returned to gain strength with J. Dalton, where these have been used to explain chemical reactions and its basic laws \cite{ref9}.

At the end of the nineteenth century, with the discovery of cathodic rays and the determination of its nature as constituted by negative electric charges and the observations about the radiation emitted by certain atoms by Becquerel and Curie have led to a new formulation of the matter constitution. The fact that atoms of a chemical element suffer a transmutation of atoms from another element through the issuance of particles with negative or positive charge, led to conjecture that the atoms should be made of these two types of particles, being the negative charges the electrons  \cite{ref10}. 

Experiences with X-ray and atoms interaction by C.G. Barkla ($1909$) led to the conclusive confirmation of electrons existing inside atoms, where those were numerically in the order of half the atomic weight of the atom ($\approx A=2$). Assumptions about the nature of positive charges could not be made due to difficulties in analyzing the properties of these particles in the gas discharge tubes, through the issue arising from radioactive substances, which did not have the uniformity of the negative charges \cite{ref10}.

Assuming that the atoms are seen in nature as electrically neutral, it was expected that the amount of positive charge was identical to the number of electrons in the atom. Another point observed is that the electrons masses, determined by R.A. Millikan in 1908(me = 0; 9108:10??27g ), is much smaller than the mass of the Hydrogen atom (the lightest of nature), which leads to the conclusion that most of the atom mass is on the positive charges, later identified as protons \cite{ref11}. The atom was not indivisible anymore, now had parts electrically active, leading to the development of atomic models in order to explain the atomic stability of which stood out the model of J.J. Thonson and E. Rutherford.
 
J.J. Thonson proposed a model in which the electrons were uniformly distributed in a positively charged sphere, with a radius around $10^{-8}$ cm, 	
which at the time was taken as the atom standard size. This model became known as the plum pudding. The collapse of this model is associated with the theoretical results expected for the scattering of particles by a few atoms, which were in stark disagreement with the experiment. This model was not able to explain the high angles of deflection caused by the particles when they pass through a thin sheet of gold.

E. Rutherford and his assistants Geiger and Mardsen did, in the year 1909, a series of experiments in which were observed large angles of scattering \cite{ref12}. The result obtained by the research group led the formulation of Rutherford`s atom model in which the atoms were seen as mini-planetary systems, with the core at the center and electrons orbiting around it. The idea was simple, the high deviations suffered by alpha particles would only be possible if the spreader center had an electric repulsive field extremely high. Since the particles had positive charge, they are doubly ionized helium atoms, the retro scattering was a result of a frontal collision with a very positive region and of high mass, which was called nuclei \cite{ref13}.
\begin{figure}[!htb]
\centering
\includegraphics[scale=0.3,width=8cm]{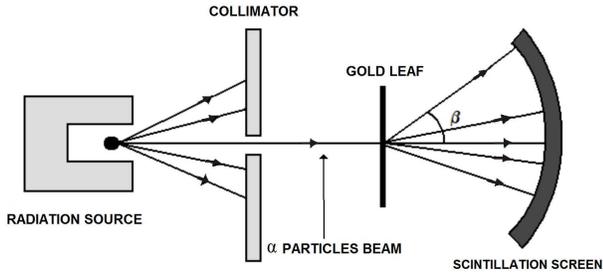}
\caption{Experiment done by Rutherford showing the high deviations ($\beta$) suffered by particles $\alpha$ when they pass through a sheet of gold and are detected in a flicker screen.} 
\label{fig1}
\end{figure}

The discovery of the nucleus was the first milestone of Nuclear Physics. The possibility of obtaining information about the atomic structure by means of collisions with sub-nuclear particles opened a range of possibilities in the search for understanding and applying energy involved in the reactions with baryonic matter, as well as knowledge on the core internal structures. Doors were being open to a new area of knowledge in which the experimental results could only be achieved by collision mechanisms. With the theories of Quantum Mechanics of W. Heisenberg and E. Schrödinger and the relativistic version of P. Dirac, a more precise theoretical description about nuclear processes can be obtained, resulting in the most modern theories of Quantum Chromodynamics D. Politzer, F. Wilczek and D. Gross and electroweak theory of S. Weinberg, A. Salam and S. Glashow, which unifies the electromagnetic interaction with the Fermi theory for weak interaction through the action of fields filled by vector bosons.

\section{NUCLEAR PHENOMENOLOGY}

The nuclear dynamics is a complex mechanism, where all the description is related to the laws of conservation, of symmetries and to the Quantum scattering formalism, besides the fact that many of the theoretical models proposed to explain the behavior of the nucleus are structured into experimental basis. Next, we will present a brief and qualitative approach of a limited number of phenomena that were chosen because of its unquestionable importance, either from a historical perspective, either from the technical sophistication and practical applicability.

\subsection{Alpha decay and Nuclear Transmutation}

We will begin with the alpha decay process. This process is directly linked to radioactive transitions experienced by unstable nucleus, where occurs the emission of $\alpha$ particle in the search for stability. The nature of these particles has been demonstrated previously, our interest in this process of decay is related to the effect of quantum tunneling or transmission through the potential barriers. This is an extremely complex process from the point of view that accurate results are related to quantum probabilities, specifically with the probability that the alpha particle is formed in the core plus the probability of it crossing the potential barrier of the latter \cite{ref14}. So that the process can be treated as a problem of collision, it is assumed, for calculation purposes, a foreign particle already formed which is to be attached to the core of departure. In this context, just calculate the probability of alpha particle going through the potential barrier of the nucleus. This is a classic problem of modern physics. See Fig.~2.

\begin{figure}[!htb]
\centering
\includegraphics[scale=0.3,width=5.5cm]{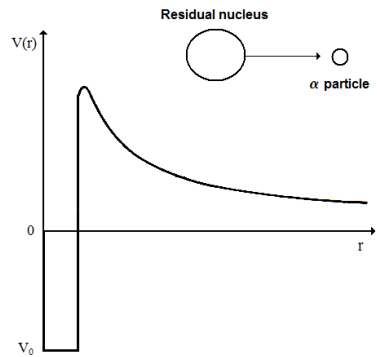}
\caption{Potential of the nucleus generated by electrostatic interaction and that must be crossed by the alpha particle.} 
\label{fig2}
\end{figure}

From the application point of view, the alpha decay served as basis for the accomplishment of the first nuclear reactions, especially in the mechanism of artificial nuclear transmutation caused by collision of cores and $\alpha$ particle. In $1919$, E. Rutherford was the first to use these particles as projectiles, at the expense of having high energy and momentum \cite{ref15}. There was however a problem, most of the $\alpha$ particles should be diverted due to the high electric field generated by the target, drastically reducing the probability of a reaction. The solution to this problem was the use of light nucleus which drastically reduces the repulsive coulomb forces increasing the probability of a possible reaction \cite{ref10}. The reactions alpha-proton alpha-neutron, where we have the emission, by the residual nucleus, from a proton and a neutron respectively, are the most significant examples of this mechanism. In the first reaction we had a nucleus of nitrogen atom being bombarded by an $\alpha$ particle according to the equation:
\be\label{equ1}
{}_{7}\mathrm{N}^{14}+{}_{2}\mathrm{He}^{4}\rightarrow[{}_{9}\mathrm{F}^{18}]\rightarrow{}_{8}\mathrm{O}^{17}+{}_{1}\mathrm{H}^{1}
\ee
The elements on the left represent the reagents, the ones on the right the compound nucleus followed by decay to the proton and residual nucleus.

The alpha-neutron reaction is very important historically, because it brought the discovery of the neutron, what led to the acceptance of a core composed by protons and neutrons, abandoning the old model that postulates the existence of electrons inside the nucleus, since this model had a number of limitations, starting with the uncertainty principle which showed that an electron confined in a region in the order of the nuclear diameter should have an energy around $100$MeV with the observed experimental value of energy for electrons in beta decay was $1$MeV. Besides the magnetic dipole moment of the core is three times smaller than the dipole moment of an electron \cite{ref11}. The reaction is based on the collision of $\alpha$ particle with a beryllium nucleus:
\be\label{equ2}
{}_{4}\mathrm{Be}^{9}+{}_{2}\mathrm{He}^{4}\rightarrow[{}_{6}\mathrm{C}^{13}]\rightarrow{}_{6}\mathrm{C}^{12}+{}_{0}\mathrm{n}^{1}
\ee
The result of this reaction provides as residual nucleus carbon $12$ and a neutron.

Both reactions can be explained from the classical theory of inelastic collisions with the increase, however, of the rest energy for each reagent given by Einstein equation $E= m_{0}c^{2}$, since the levels of energy involved still do not require a deeper relativistic treatment. The process of transmutation is very useful because it allows the creation of artificial compounds, "eliminating", in parts, the need for exploitation of natural resources.

Another interesting application of the alpha decay can be observed within the smoke detectors (fire alarm), where is used the radioactive element called americium- $241$, which has half life of $458$ years. Smoke detector devices are low cost and are widely used in commercial and residential buildings. It is composed of two parts, a sensor to sense the smoke and an electronic campaign. There are two types of detectors: the photoelectric and the ionization ones. For practical purposes, within the context of this work, we will stick to the ionization detectors. Within these detectors we have an ionization chamber which has in its interior $2\times10^{-4}g$ of Americium-$241$, which equals $0,9$microcurie, this amount of americium suffers approximately $37$ thousand nuclear transmutations per second and in each is emitted an $\alpha$ particle. The ionization chamber consists in two plates subject to a potential difference, as can be seen in Fig.~3

\begin{figure}[!htb]
\centering
\includegraphics[scale=0.3,width=7cm]{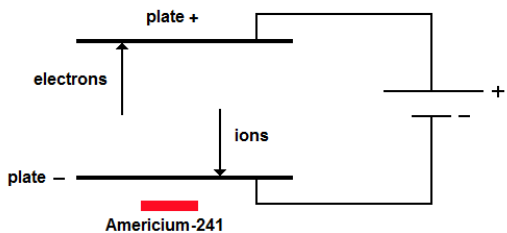}
\caption{Schedule of ionization chamber operation of a smoke detector.} 
\label{figb1}
\end{figure}
The $\alpha$ particle ionize the atoms of oxygen and nitrogen from the air inside the chamber. The electrons released in this process are attracted to the positively charged plate and the positive atoms are attracted to the plate with negative charge, establishing therefore an electric current between two plates originated from the electrons and ions movement. In a situation of fire, the smoke particles enter the ionization chamber, react with the ions, making them neutrons, interrupting the current between the plates. The detector senses the drop in current and triggers the alarm.

\subsection{Beta decay}

In $1914$, J. Chadwick was the first to observe experimentally, through measurements performed with magnetic spectrometers, that the nucleus could emit electrons \cite{ref14}. These initial observations led to believe that the electrons were, such as protons, the constituents of the nucleus, which was later refuted by the discovery of the neutron. This, such as the alpha decay,  is about a procedure of radioactive transition between states of some unstable nuclei with the emission of high energy electrons, which was called beta decay. The original theory of beta decay had serious problems because it could not handle the energy spectrum observed experimentally and could not be committed by a single electron. In $1930$, W. Pauli postulated the existence of the neutrino, a particle that was also emitted in decay, without charge and mass and spin $1/2$; the existence of a particle devoid of mass, with null charge and spin momentum $1/2$ was necessary to preserve the principles of energy and angular momentum conservation. A more precise theory was only proposed in $1934$ by E. Fermi \cite{ref16}, in which concludes a new type of interaction, the weak interaction. 

Years later this theory was enhanced with the work of R. Feynman and M. Gell-Mann \cite{ref17}, where a consistent relativistic treatment from the Dirac equation was used to describe the Fermi interaction. Inside the nucleus the beta decay can be expressed by the following reactions:
\begin{align*}
\mathrm{p}\rightarrow\mathrm{n}+\mathrm{e}^{+}+\nu\tag{$3.a$}\\
\mathrm{n}\rightarrow\mathrm{p}+\mathrm{e}^{-}+\bar{\nu}\tag{$3.b$}
\end{align*}  
The equations show the positive and negative beta decay with the emission of a positron and a neutrino or the emission of an electron and an anti-neutrino, respectively.
\begin{figure}[!htb]
\centering
\includegraphics[scale=0.3,width=6.5cm]{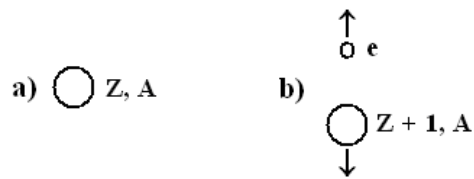}
\caption{beta decay(Z$\rightarrow$ atomic number and A$\rightarrow$ number of mass): a) initial state of the nucleus before the decay. b) Nucleus after the emission of electron.} 
\label{fig3}
\end{figure}

The applications of this decay mechanism are present in a class of important phenomena, which extend from the human physiology, in the medicine, to the development of space technology and industrial technology. In the last case, we have the use of Promethium, the chemical element of atomic number $61$, which is found at room temperature at solid state. It is used as an emitter of beta particles in the construction of thick meter, within the metrology of precision in the construction of the clock dials and pointers. In the aerospace industry is used to manufacture micro batteries for long periods and possibly as a portable source of X-rays and heat in space probes and artificial satellites \cite{ref18}. 
	
An interesting phenomenon that involves the beta emission occurs inside the human body, this radiation procedure is responsible for the functioning of the heart muscle. In the human blood, is diluted about $30$mg of potassium-$40$ ($K^(40)$), which is a radioactive isotope of potassium-$39$, this concentration is enough to the ejection of beta particles. Within the heart, there are two cavities, called left and right auricle, the first has the function of pumping blood into the pulmonary circulation and the second is responsible for distributing blood enriched with oxygen to all parts of the body, are endowed with cells and beam of nerve fibers called the sino-atrial node and Bachman beam respectively. The potassium diluted in the blood, inside these cavities, suffers disintegration by electrons emission (beta particles), when these electrons collide with the nerve terminations of the heart cavities walls, generate a stimulus that leads the heart muscle to contract and grow in a determined rhythm, so that the blood is pumped to all parts of the body.
 
\subsection{Möessbauer Effect}

Following the line of reaction induced by gamma rays, another mechanism gets prominent called absorption by nuclear resonance or Möessbauer effect. The description is as follows: after absorbing a photon, the nucleus gets to the excited state, the line width associated with this state is given, according to the uncertainty principle, by $\Gamma=\hbar/\tau$, where $\tau$ represents the average life of the state. After this period, the nucleus decays by emission of a new $\gamma$ photon, suffering a retreat. The central idea is that the energy transferred to the core of the absorption must be equal to the recoil energy in the issue, forecasting a resonant state.

Till now, we have considered the core of a free atom, the problem changes configuration  for the case in which the atom is connected to a crystal network. Due to the connections with the network, the core can absorb a photon without retreat, it says that resonant absorption without recoil occurred, what is precisely the Möessbauer effect. The fact that the atom can perform harmonic vibrations along three freedom degrees, where the energy associated with the oscillation modes is quantized and given by $\epsilon = \hbar\omega$ , where $\omega$ is the network oscillation frequency and the value of $\epsilon$ represents the vibrations quantum called phonon, making that, in the ejection process, the recoil energy get transferred to the network in the form of oscillations, contributing to temperature raising of the same, given that this phenomenon only occurs if the recoil energy is greater than the energy of the net quantum oscillation, generating a state of excitement in it. If the recoil energy is less than $\hbar\omega$, the network can not be excited, performing as a buffer; occurs, therefore, an ejection without recoil. This effect is widely used in astrophysics, the lines of the Möessbauer effect spectrum provides a precise characterization of the elements that compose the interstellar matter \cite{ref14}.

Other applications of Möessbauer spectroscopy lie in structural characterization of some proteins, in the chemical composition of meteorites. In industry, can be used to study the structure of alloys and geology for geological dating of ceramics \cite{ref19,ref20,ref21}.

\subsection{Strong Nuclear Force}

As major contribution of the collision mechanisms, the nuclear physics is rooted in the characterization of the strong nuclear force; this is the force responsible for connecting protons and neutrons inside the atom nucleus and, consequently, ensure the stability of the core and the hadronic matter; is the force of greater intensity in nature, which justifies calling it strong interaction. The hadronic interaction can be described from the neutron-proton dispersion at high energy. This process is described in terms of the bounded energy of the deuteron, which is a nucleus composed of one proton and one neutron. The problem is physically treated with quantum principles of interaction between two bodies. 

The results of scattering experiments between nucleons show the following properties about the nucleon interaction: $1$) the property of saturation, ie a single nucleon can only interact with a restricted number of nucleons; $2$) the independence of the charge, ie the nuclear forces are symmetrical in relation to the charge, the intensity of interaction a reaction ($\mathrm{nn}$) is identical to a reaction ($\mathrm{pp}$); $3$) the presence of exchange forces. In quantum mechanics, when two particles interact mutually, there is always a state in which one property can be shared, producing an exchange interaction between to the nucleons. Heisenberg supposed that this property would be the charge \cite{ref24}.

The property of energy exchange is closely linked to the mesonic theory of nuclear forces, proposed by the Japanese physicist H. Yukawa in $1935$ that, in analogy with the electromagnetic interaction mediated by the exchange of photons, assumed that the hadronic interaction between nucleons would be result of the exchange of one meson $\pi$. The fact that this particle have mass, allowed the description of this interaction as short-range. Moreover, the mesons have charge ($\pi^{-}$, $\pi^{+}$) or may be electrically neutral  ($\pi^{0}$), needed requirement to, in an interaction, the involved particles charge get exchanged. Explicitly in this process, a proton can become a neutron and vice versa. See Fig.~5.
\begin{figure}[!htb]
\centering
\includegraphics[scale=0.3,width=4.5cm]{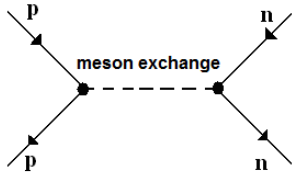}
\caption{Feynman diagram for illustrating the interaction between a proton and a neutron by exchanging a meson.} 
\label{fig4}
\end{figure}

The greater application of the strong nuclear force occurs in the nuclear fission reactors, used to produce electricity; where the uranium, when bombarded by neutrons, absorb these particles and become unstable, suffering a split (creating two new lighter atoms). The energy corresponding to the strong nuclear force that united the protons and neutrons in the nucleus of uranium is released as kinetic energy from both residual nuclei. This energy can be used within the reactor of a nuclear plant or can be used for war purposes, such as the atomic bomb, with devastating effects.

\subsection{Nuclear fission}

There are two reaction mechanisms in Nuclear Physics that are intrinsically linked, they are: the moderation of neutrons and nuclear fission. In $1938$ O. Hahn and F. Strassmann \cite{ref23} discovered nuclear fission; Liese Meitner and R. O. Frisch \cite{ref24} interpreted the first time the mechanism of the process and then N. Bohr and J. Wheeler \cite{ref25} proposed a theoretical treatment based on the model of the liquid drop. Finally, only in 1942 E. Fermi, manage the first controlled chain reaction \cite{ref14}.

The process of nuclear fission is the result of a dynamic instability of the nucleus, which results in its splitting into two residual nuclei with a high release of energy. For nuclei with $Z > 92$, there is the possibility of spontaneous fission to occur without an external agent acting \cite{ref23}. 	However, for practical purposes, the process of fission induced in nuclei with $Z\approx 90$ (uranium 233, for example) generated in nuclear reactors, as in Fig.~6, presents bigger efficiency. The mechanism is based on the capture of a neutron by a uranium nucleus through a collision, although there is the possibility to promote the fission of a uranium nucleus through collisions with high-energy particles, protons, deuterons and $\gamma$ rays \cite{ref24}.

\begin{figure}[!htb]
\centering
\includegraphics[scale=0.3,width=9cm]{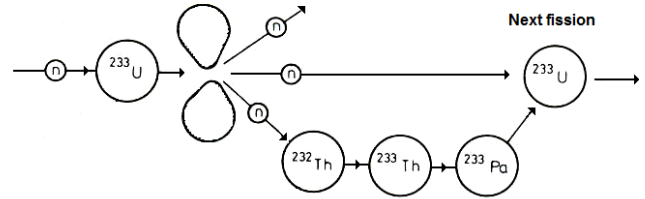}
\caption{Schedule of nuclear fission in a reactor where the core release neutrons that can be captured by other non fissile that after fell to $U^(233)$, or are absorbed by fissionable material and induce fission.} 
\label{fig6a}
\end{figure}

The neutrons energy generated in nuclear reactions and in the fission process is in the order of $1$MeV \cite{ref23}, ie, they are extremely fast, what makes hard its capture by other nuclei of uranium involved in the reaction. To get a chain reaction, the neutrons must be thermalized, or moderate, from the collisions with other nuclei (such as graphite, beryllium, etc.). With the decrease of its speed, increases the probability of a reaction with a uranium nucleus, which must be to fissioned, releasing neutrons that will be moderated and captured by other nuclei, starting a chain reaction. This process is used in thermonuclear reactors for power plants to produce electricity and manufacture of mass destruction weapons like the atomic bomb.

\subsection{Enrichment of Uranium}

The International Energy Overview 2007 (IEO2007), published by the International Energy Agency (IEA) estimates a growth of $57\%$ in world consumption of the various forms of energy between the period 2004 and 2030. According to the report, all energy used worldwide will grow from $447\times10^{12}$ Btu (British Thermal Unit) in 2004 to $702\times10^(12)$ Btu in 2030. For nuclear energy, the average consumption should grow $1.3\%$ in that period, with greater concentration in developing countries, whose estimated demand is of $4.2\%$.

In this scenario, the uranium enrichment plants gain importance, whose related researches worry about to make it feasible, from the economic, social, environmental points of view, among others, the processes for fuel production, from the prospecting of ore to the final transportation of radioactive waste.

Uranium found in nature, under the form of uranium dioxide ($\rm UO_2$), is composed of approximately $99.3\%$  of the isotope $^{238}\rm U$ and $0.7\%$ of $^{235}\rm U$. The latter, unlike the $^{238}\rm U$, is a fissile material, , in other words, that suffers fission only caused by low energy neutrons (thermal neutrons). The enrichment of uranium consists, therefore, in the increasing of $^{235}\rm U$ concentration that, in the fuels of the reactors, should be $2-4\%$. An atomic bomb, on the other hand, is built with uranium enriched to $95\%$.

One of the isotopes separation processes - the gaseous diffusion - consists in transforming $\rm UO_2$ into uranium hexafluoride gas ($\rm UF_6$) and to make this gas spread trough porous plates, separating the ($^{235}\rm UF_6$) of the ($^{238}\rm UF_6$). This procedure was used in the United States during the last world war, for large scale production of highly enriched uranium. Another process that have industrial prominence is the ultracentrifugation, in which do to rotate the gas $\rm UF_6$ in a cylinder at high speed; as a result, the $^{238}\rm UF_6$ is concentrated on the cylinder wall, while the $^{235}\rm UF_6$) remains in the center \cite{ref22,ref28b}.

\subsection{Nuclear Reactors}

Nuclear reactors are systems in which, controlled, nuclear reactions are produced in chain to release energy. The physics of reactors basically study the phenomena related to the behavior of neutrons in a set of special medium materials, arranged in appropriate quantities and geometries. The reactors are classified by several factors: building materials, fuel, geometry, purpose of use, among others. We can, however, for simplicity, share them in two groups: thermal reactors, that use moderators to reduce the energy (speed) of neutrons and rapid reactors, using neutrons without moderate them \cite{ref22,ref28b}.

The essential parts in a reactor are: an active core with the fuel element - usually uranium - where the fission reaction is kept, A moderator - water or graphite - to reduce the energy of neutrons, a reflector to avoid its exhaust, a cooler to remove the heat generated in consequence of the fissions, and a shield to block the passage of penetrating radiation \cite{ref1h}.

We treated the thermal fission reactor particularly, whose purpose can be lead the reaction of nuclear fission for use in research or for conversion and use of energy in new form (reactor power). The function of a reactor in a power system is to convert energy from fission to thermal energy, preparing the subsequent conversion of thermal energy into electrical energy. The fission energy is converted into kinetic energy of the fission fragments, and the immediate result is the increase in the internal energy of the material fuel and moderator. In a nuclear plant, the increase of internal energy generates steam, which triggers a turbine (outside the reactor), making an electric generator \cite{ref2h}, as seen in Fig.~7. 

\begin{figure}[!htb]
\centering
\includegraphics[scale=0.3,width=7cm]{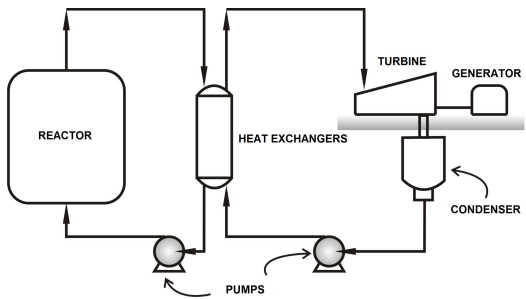}
\caption{Simplified scheme of a Thermonuclear Power Plant.} 
\label{fig6}
\end{figure}

A fission produced by a neutron releases new neutrons capable of producing new fissions. For each neutron absorbed, about 2:5 new neutrons are released. It is necessary then that at least one of neutrons released cause a new fission, to establish the desired chain reaction. The rate of reaction is controlled inserting or removing (bars control) - of boron or cadmium - which absorb neutrons without suffering further reactions. These bars are also set as security devices.

\subsection{Nuclear fusion}

The last case to be addressed is nuclear fusion. We have seen previously that heavy cores could fission by a collision with neutrons, liberating a great deal of energy. The mechanism of fusion is almost the reverse: fast and light nucleus can collide and merge to form heavier nuclei, where a considerable amount of energy is released in this process. 

This energy is associated with dissipation of heat, depends directly on the masses of the partners involved in the reaction and their properties are related to nuclear matter, ie, for fusion to occur some requirements must be fulfilled by the partners involved in the process: 
$1)$ the kinetic energy of the nuclei reaction must be large to allow the increase of penetration probability in Coulomb barrier, this process occurs in very light nuclei at a temperature of a 107K, where the atoms are fully ionized foreshadowing a state of plasma \cite{ref14}; $2)$ the matter density in the temperatures involved in the fusion reaction, must be extremely high.

The interior of the stars, in particular the sun, has a whole propitious scenario to this type of reaction, the density of the sun interior is approximately $1000g/cm^{3}$ at a temperature of $1,5.10^{7}K$. The Fig.~8 represents the fusion reaction of hydrogen helium that occurs inside the stars and that was present at the beginning of the formation of the universe in primary nucleosynthesis \cite{ref28}.
\begin{align*}
\mathrm{H}^{1}+\mathrm{H}^{1} &\rightarrow\mathrm{H}^{2}+\mathrm{e}^{+}+\nu+0,42\mathrm{MeV}\tag{$4.a$}\\
\mathrm{H}^{2}+\mathrm{H}^{1} &\rightarrow\mathrm{He}^{3}+\gamma+5,49\mathrm{MeV}\tag{$4.b$}\\
\mathrm{He}^{3}+\mathrm{He}^{3} &\rightarrow\mathrm{He}^{4}+2\mathrm{H}^{1}+\gamma+12,86\mathrm{MeV}\tag{$4.c$}
\end{align*}  

\begin{figure}[!htb]
\centering
\includegraphics[scale=0.3,width=5.5cm]{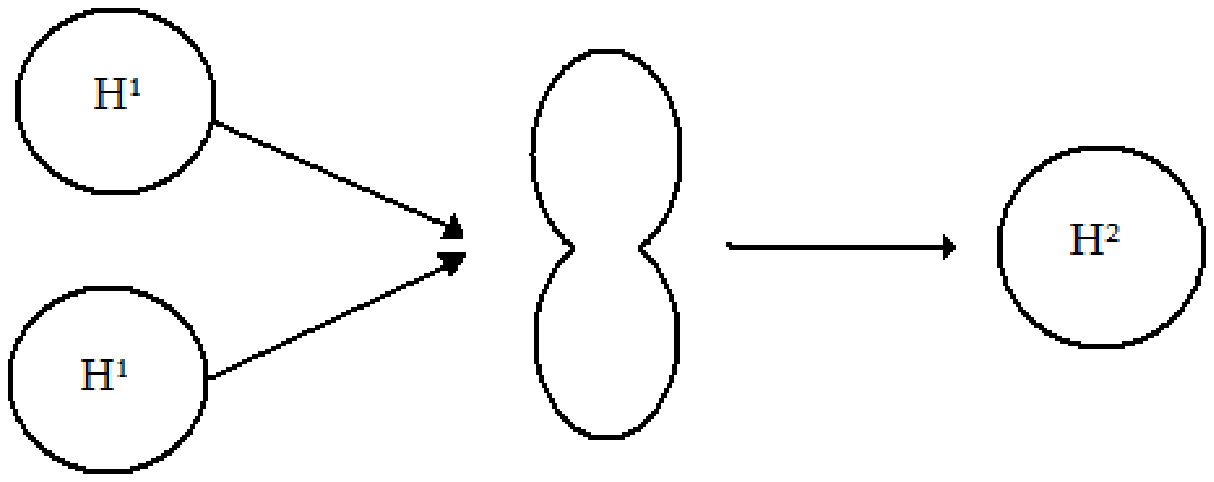}
\caption{Schematic of reaction of nuclear fusion in which two protons merge to generate the deuteron.} 
\label{fig7}
\end{figure}

According temperature, heavier nuclei can be formed. The major application of nuclear fusion is related to generation of electrical energy to replace the nuclear fission, but in a more clean and safe way.

The main advantages related to the current fission reactors are: 1) Easy to obtain fuel and in great quantity, the deuterium can be obtained from sea water and tritium obtained in own reactor fusion from lithium, the uranium used in fission is very rare and difficult to extract. 2) the fusion process is safer than the fission process, since the amount of fuel used is lower, without uncontrolled liberation of energy. Besides, the radiation taxes emitted are inferior the tax of natural radiation that happens in the terrestrial surface. 3) less production of nuclear waste compared to fission, in addition to the fact that the waste coming from fusion is not raw material to build nuclear weapons, like in fission case. 

Currently, NASA has invested in research on fusion nuclear reactors construction to generate energy for space rockets. Fusion propellants would be more efficient and make the rockets faster in addition to provide longer trips, because the fuel (hydrogen) would be unlimited generated.

\section{CONCLUSION}

The most modern strands of education promote a teaching focused on building a dynamic and constructive mentality on student. Based on this aspect, this work explored the possibility of themes that show  too complex mathematical formalism in Nuclear Physics, to be worked at high school. The conceptual and philosophical structure of the phenomena, under certain aspect, can be maintained without appeal to mathematical formulae, exposing practical applications that can be identified by the student in the scientific-economical actual scenario. In that point, the intervention of the educator is made necessary as a bridge between theoretical knowledge and practical contextual knowledge. Remains as suggestion for educators the use of didactic resources as animation and videos, very disseminated on the internet nowadays, which handle these matters. In general, science should be disseminated among the young in order to awaken them to the reality of the technological society, serving as a vehicle for inclusion and as a stimulus to create a new generation of scientists.

\vspace{0.4cm}
\noindent{{\bf Acknowledgments}}

M.A.M Souza thanks the Breno Oliveira and Janaina Camargo by discussions and Joselma Rangel by the aid with the figures.

\vspace{0.3cm}

\end{document}